\documentclass[floatfix,pra,superscriptaddress,reprint,longbibliography]{revtex4-1}
\usepackage{graphicx}
\usepackage[caption=false]{subfig}
\begin{document}

\title{Training the trainer: Professional Development for High School Physics Teachers with Low Physics Background}

\date{\today}

\author{Anne Wang}
\affiliation{Department of Physics and Astronomy, Texas A\&M University,
		College Station, TX~~77843}
\email[send correspondence to: ]{etanya@tamu.edu}
\author{Jonathan Perry}
\affiliation{Department of Physics, University of Texas,
		Austin, TX~~78712}
\author{Matthew Dew}
\affiliation{Department of Physics and Astronomy, Texas A\&M University,
		College Station, TX~~77843}
\author{Mary Head}
\affiliation{Lamar Consolidated Independent School District, Rosenberg, TX~~77471}
\author{Tatiana Erukhimova}
\affiliation{Department of Physics and Astronomy, Texas A\&M University,
		College Station, TX~~77843}

\begin{abstract}
The shortage of highly qualified high school physics teachers is a national problem. The Mitchell Institute Physics Enhancement Program (MIPEP) is a two-week professional development program for in-service high school physics teachers with a limited background in the subject area. MIPEP, which started in 2012, includes intense training in both subject matter and research-based instructional strategies. Content and materials used in the program fulfill state curriculum requirements. The MIPEP curriculum is taught by Texas A\&M University faculty from the Department of Physics \& Astronomy along with two master high school physics teachers. In this paper we present the design and implementation of MIPEP. We report on assessment of knowledge and confidence of 2014-2018 MIPEP cohorts. We also present the results of the 2020 program that was delivered remotely due to the pandemic. Analysis of these assessments showed that the majority of MIPEP participants increased their physics knowledge and their confidence in that knowledge during both traditional and virtual program deliveries. 
\end{abstract}

\maketitle

\section{Introduction}
The past 30 years have seen a significant increase in the number of high school students taking physics, to almost 1.4 million students in 2013 \cite{White&Tesfaye2014}. Given efforts to boost STEM interest and enrollment in the US, it is unlikely that this number has plateaued in the past several years. Prior literature indicates the importance of pre-college physics courses to student success in STEM careers. Students who took two years of high-school physics have better chances of doing well in gateway introductory physics classes \cite{Sadler2001}. Taking quantitative physics in high school increases STEM career interest and chances to obtain a STEM degree in college \cite{Tyson2007, Tyson2011, Sadler2014}. 

An overall unpreparedness of high school physics teachers has long been a concern \cite{McDermott1975, Wheeler1998, Hodapp2009, White&Tesfaye2012, Rushton2017}. In their 2009 paper published in Physics Today, Hodapp et al. indicated that more than 23,000 US high-school physics teachers are not adequately prepared to teach the subject; with only one third of them having majored in physics or physics education \cite{Hodapp2009}. The authors went on to conclude that ``poor teacher preparation denies students access to a quality education in the physical sciences." The results of a 2012-2013 Nationwide Survey of High School Physics Teachers \cite{White&Tyler2014} showed that less than half of high school physics teachers have a background in physics implying that the majority of physics teachers possess only a basic knowledge of the subject at best. In another study, Banilower et al. found that more than a quarter of all high school physics teachers have not taken any course in the discipline beyond the introductory level, contributing to physics teachers feeling less well prepared to teach their subject than other science teachers \cite{Banilower2013}. Also, about 40\% of high school physics teachers teach a majority of their classes in a subject other than physics \cite{White&Tyler2015}. This lack of preparation for US high school physics teachers has been in parallel with an increasing enrollment in physics courses, where the American Institute of Physics (AIP) showed that 39\% of high school students who graduated in 2013 took at least one physics class \cite{White&Tesfaye2014}. An increase in the number of high school students taking physics can only create a greater demand for qualified physics teachers. Currently, the rate of preparation of high school physics instructors is insufficient, and a growing number of instructors in the field are crossover teachers who have no substantial formal training in either physics or physics teaching \cite{Meltzer2012}. This has resulted in a growing number of students who take physics from teachers with inadequate preparation, according to the Phys21 report by the American Physical Society (APS), American Association of Physics Teachers (AAPT), and AIP \cite{heron_mcneil_2016}.

Recent studies indicate that students who took physics classes with a certified physics teacher performed better than their counterparts taking physics with an out-of-field teacher \cite{Sheppard2020, Krakehl2020}. Further, after analyzing data from New York State public schools with more than 1,500 physics teachers and 47,000 physics students, Krakehl et al. \cite{Krakehl2020} raised a concern that 40\% of physics teachers were isolated, and their students tended to have weaker physics performance scores than students of non-isolated teachers. Survey data from 2012 suggests this isolation may be much higher across the country, where there is only one physics teacher in over 80\% of schools \cite{White&Tesfaye2012}. Such isolation can lead to occupational stress and teacher attrition, especially in early career teachers \cite{Dussault1999, Borman2008, Beltman2011, Buchanan2013, Schlichte2005}. This could be especially true for a state as large as Texas. School-level socioeconomic status has also been shown to be a main negative predictor of student physics performance indicating that the shortage of qualified physics teachers disproportionately affects students from economically disadvantaged school districts and may effectively deny them opportunities to pursue STEM-related degrees and occupations \cite{Kelly2009}.

According to literature, it is essential for teachers to develop physics pedagogical content knowledge, which is the combination of subject knowledge and instructional strategies that is unique to teachers within the field \cite{Etkina2010, Magnusson1999}. To develop this personal form of understanding, teachers must be well versed in the content of their field \cite{Shulman1987}. Physics in particular requires a depth of knowledge that is best gained through interactions with a university or college physics department \cite{McDermott1990, Foote2018}. This subject specific expertise is particularly vital in helping students overcome their misconceptions \cite{Etkina2010}.

Some existing programs are working to meet demands for well-trained high school physics teachers. Programs focused on pre-service teacher training exist at many universities and have the goal to increase the graduate rate of majors intending to teach at the high school level \cite{Scherr2014}. These programs provide a solid path to improve the percentage of high school physics teachers with a background in the discipline. However, according to data from PhysTEC, such programs are not currently close to meeting needs in any state \cite{phystec}, and they leave a gap for teachers already in the profession without a background in the subject. The effectiveness of these programs may be boosted in the coming years through the efforts of programs such as \textit{Get the Facts Out} \cite{adams2019}. Other programs including AP Summer Institutes and the University of Texas OnRamps dual enrollment program do offer professional development for in-service high school physics teachers, though both are focused on serving a particular curriculum associated with their programs \cite{APSummerInstitutes, OnRamps}. This leaves a niche for professional development programs looking to improve physics understanding for in-service teachers, such as the Texas A\&M University Mitchell Institute Physics Enhancement Program (MIPEP) \cite{mipep}.

MIPEP was developed as a pilot program to help in-service high school physics teachers who lack a substantial (or any) background in physics. The program seeks to develop a deeper content knowledge and combine that with research-based instructional practices for the physics classroom. This work presents the design, implementation, and an assessment of the MIPEP program. We evaluated gains in participant physics content knowledge, their confidence in that knowledge, and retention for cohorts between 2014-2018. We also present the results of the 2020 program that underwent an emergency redesign and was delivered remotely due to the pandemic. This work may serve as a model or instructive example to other institutions considering developing a content-focused professional development program for in-service high school teachers.

\section{Program}
 As a professional learning opportunity, the MIPEP curriculum and experience was designed to meet the following learning goals:
\begin{itemize}
 \item positively impact physics teaching and learning
 \item increase participants' physics content knowledge
 \item assist participants to develop and use research-based instructional strategies 
 \item provide laboratory-based learning experiences
 \item encourage collaboration among physics educators.
\end{itemize}

First offered in summer 2012, MIPEP is a two-week intensive professional development program that targets in-service Texas high school physics teachers with little or no background in the subject. Physics content aligns with the Texas Essential Knowledge and Skills (TEKS) criteria, which is established by the Texas Education Agency. Content focused instruction is facilitated by faculty from the Department of Physics \& Astronomy contributing 66 contact hours to the program. Faculty focus primarily on clarifying fundamental concepts typically found in a two-semester introductory physics sequence, Table \ref{table:1}, at the level of algebra-based instruction paired with problem-solving and hands-on demonstrations. The program also includes two days of laboratory activities and more than five hours engaging with hands-on physics experiments selected from hundreds available from the department. Additional programmatic components include pedagogical development, extracurricular opportunities, and administrative tasks including pre- and post-assessments.

Day long content-oriented sessions led by physics faculty are followed by late afternoon and evening sessions led by two master teachers who focus on pedagogy and pedagogical content knowledge. These sessions combine physics knowledge with research-based instructional strategies, showcasing different activities that teachers can use in their classrooms. As an example, after covering two-dimensional motion, participants will determine when to release an object from a second floor balcony such that it will hit an object moving in a straight line across the floor below. The object moving can be a simple system, such as a block sliding across the floor, or a more complex system, such as a bowling ball being pushed with a broom. Including master teachers as part of the MIPEP team brings essential contact with experienced physics teachers to assist participants in their individual growth as an educator.

\begin{figure}
\centering
\includegraphics[width=8.6cm,height=6.88cm]{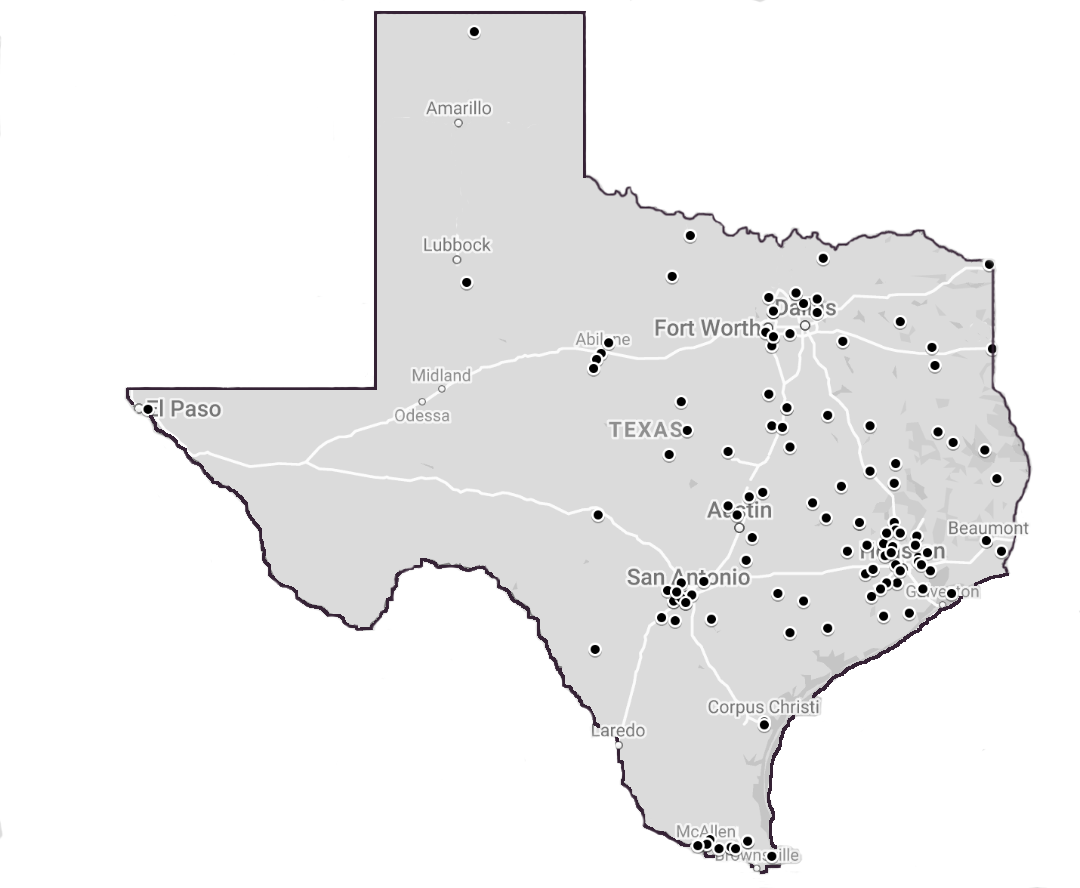}
\caption{\label{Zipcodemap} Participant geographical demography for 2015-2018, and 2020.}
\end{figure}

For the first two years of the program, teachers were hosted at a private facility of a major donor. Starting in 2014, the program moved to Texas A\&M University. The primary support for MIPEP has been provided by funding from the Cynthia and George Mitchell Foundation. Participants receive campus lodging, meals, and travel reimbursement in addition to all program events and materials. Since its inception, MIPEP has increased its annual cohort from 15 participants from 12 districts in 2012, to 26 participants from 24 districts in 2019. As seen in Figure \ref{Zipcodemap}, participants come from all over the state, as well as from a mixture of urban, suburban, and rural districts. More than half of participants from 2014-2018 were female, with about 70\% identifying as white, and 18\% identifying as Hispanic, Table \ref{table:2}. These numbers are similar to teacher demographics in Texas \cite{Ramsay2019}. Many participants have reported completing an alternative certification program to receive their Texas teaching credentials. Applicants to MIPEP may be any Texas high school teacher who has taught or will be assigned to teach physics courses. The program specifically selects applicants with little to no physics background, which is defined as 0-2 college-level physics courses in an applicants' educational background. Teachers interested in the program submit a letter from their principal confirming that they are assigned to teach physics in the next school year, and priority is given to applicants with little to no background in the field. A sense of community is fostered among teachers through frequent group activities throughout the program, and all participants are added to a MIPEP listserv to facilitate continuing interactions after the program. Recruitment and program advertising is done through Texas science teacher conferences, emails to school districts, and word of mouth.

Upon completion of the program, MIPEP participants receive Continuing Professional Education and Gifted/Talented Credits. The teachers who completed the program received 80 credit hours of Continuing Professional Education, including 34 credit hours of Gifted \& Talented curriculum and 6 credit hours of Gifted \& Talented annual update. In addition to rigorous training in the high school physics curriculum, teachers are exposed to ongoing, significant physics research through a variety of enrichment activities which are guided by faculty. These include hands-on demonstrations and ideas for low budget experiments, lunches with faculty, and tours of local research facilities, including Texas A\&M's Cyclotron Institute, Nuclear Reactor Laboratory, and Observatory. 

In June 2020, MIPEP was reformatted to an online course as a result of the emergency switch to remote teaching and learning. The classes were taught synchronously via Zoom. In-class demonstrations were pre-recorded and showed as videos. Shoe boxes with equipment for low budget demonstrations were assembled and mailed to each teacher. Evening discussions were led by master teachers who showed live-streamed demonstrations and shared tips on engaging students. Due to the shift to an online program, no labs or facility tours were included. The program was conducted over a two-week period similar to past years. Lectures by an astronaut and prominent physicists were delivered via Zoom. Due to the program being offered virtually, MIPEP 2020 was able to accept more participants compared to previous years (40 teachers registered) since there were no living expenses and online delivery of the course reduced the barriers for teachers to leave their families. Expenses for MIPEP 2020 were reduced by about 2/3 compared to the previous year, with approximately 1/3 of the remaining budget applied towards teacher supplies.

\begin{table*}
\centering
\begin{tabular}{ |p{6cm}|p{2cm}|p{6cm}|p{2cm}|  }
 \hline
 \multicolumn{2}{|c|}{Week 1} & \multicolumn{2}{|c|}{Week 2}  \\
 \hline
Topics Covered & Assessment & Topics Covered & Assessment \\
& Item & & Item \\
 \hline
 Kinematics, Vectors and Graph Analysis   & 1-11, 14   & Oscillations and Waves & 32, 36 \\
 Newton's Laws  &  15-20, 27 & Electrostatics   & 21, 23, 25, 27\\
 Work, Power, Energy, and Conservation of Energy   & 51, 53-59, 61 & Magnetic Fields, Magnetic Forces, and Electromagnetic Induction & 24, 26, 27, 44-46 \\
 Momentum, Impulse and Conservation & 28-30  & Currents and Circuits (R,C)   & 47, 48, 60-62 \\
 Rotational Motion: Kinematics and Dynamics & 12, 13, 49, 50, 52 & Gravity and Law of Universal Gravitation & 20, 22  \\ 
  &   & Atomic, Nuclear and Quantum Physics & 39-41 \\
  &   & Electromagnetic Waves and Optics & 31, 33-35, 37, 38, 42, 43 \\
 \hline
\end{tabular}
 \caption{Topics covered during MIPEP and their corresponding assessment numbers.}
\label{table:1}
\end{table*}

\begin{table*}
\centering
\begin{tabular}{lcccccc}
 \hline
Characteristics & 2014 & 2015 & 2016 & 2017 & 2018 & 2020 \\
 \hline
Female	& 9 & 12 & 14 & 19 & 12 & 18 \\
Male & 8 & 6 & 4 & 4 & 12 & 22\\
\hline
White, not of Hispanic descent & 10 & 15 & 12 & 15 & 18 & 32\\
African American or Black & 0 & 0 & 2 & 4 & 0 & 1 \\
Latinx & 5 & 1 & 3 & 3 & 6 & 5\\
Asian & 1 & 2 & 0 & 2 & 0 & 2\\
Native American \& Alaskan Native & 1 & 0 & 0 & 0 & 0 & 0\\
\hline
Conceptual Physics &7 & 5 & 5 & 5 & 3 & 0\\
On-level, math-based physics & 5 & 15 & 17 & 17 & 20 & 34\\
Pre-AP Physics & 6 & 6 & 3 & 8 & 4 & 4\\
AP Physics B (1\&2) & 1 & 2 & 3 & 1 & 5 & 2 \\
AP Physics C & 4 & 0 & 0 & 0 & 0 & 0\\
\hline
\end{tabular}
\caption{\label{table:2} Self-reported demographics of participant gender, ethnicity, and courses currently scheduled to teach for MIPEP 2014-2018, and 2020. The program only allowed binary responses for gender. Participants were able to respond with a single ethnicity 2014-2016, and with multiple ethnicities thereafter. In 2016, one participant declined to identify their ethnicity.}
\end{table*}

\section{Data Collection and Analysis}
To measure the program's impact on physics teachers' content knowledge and their confidence in that knowledge, a diagnostic pre- and post-program assessment was used. A subsequent assessment was also sent to prior participants in fall 2018 to measure persistence of knowledge and confidence. This assessment was assembled by the master teachers drawing on multiple choice items from Physics Bowl questions, an annual competition for high school students run by AAPT \cite{aaptphysicsbowl}. In total, a 62-item assessment was created from topics found in a traditional two semester college or university introductory physics course sequence, Table \ref{table:1}. In addition to their content responses, teachers also indicated confidence in their answers on a 5-point Likert scale for each question.

We considered the normalized gains for both physics knowledge and confidence. The normalized gain was calculated using
\begin{equation}
g = \frac{\text{Score } 2 - \text{Score } 1}{100\%- \text{Score } 1} 
\end{equation}
where Score 1 is a participant's pre-program score, and Score 2 is either the participant's post-program score or their score on the subsequent assessment \cite{Hake1998}.

Each year the program's master teachers administered the assessment at the beginning and end of the program, providing pre- and post-program evaluation. The assessment was administered to all MIPEP participants ($N=100$) from 2014 to 2018 with $N=97$ usable responses for knowledge and $N=94$ usable responses for confidence. Assessments were typically conducted on paper and were introduced as a diagnostic tool. Data were not collected from the 2019 cohort for this study. In 2020, due to the online format, Google Forms were used to administer the same assessment. The 2020 program yielded $N=33$ usable responses of the 40 participants.

To measure the persistence of gains made during the program, the same assessment was distributed to prior participants online in fall 2018. Each participant was given one hour to complete the assessment (same as during the program). The assessment was made available to prior participants for one month. Two reminder emails were sent to encourage participation.

The subsequent assessment was sent to all 139 past program participants, including 2012 and 2013. Response to the subsequent assessment was voluntary, resulting in a 32\% response rate ($N=45$). Only responses which were complete and could be matched with previous data were used in this study. The subsequent assessment responses gave 19 usable results, involving participants from 2015-2018. 

\section{Results}
\begin{figure}
\centering
  \includegraphics[width=8.6cm]{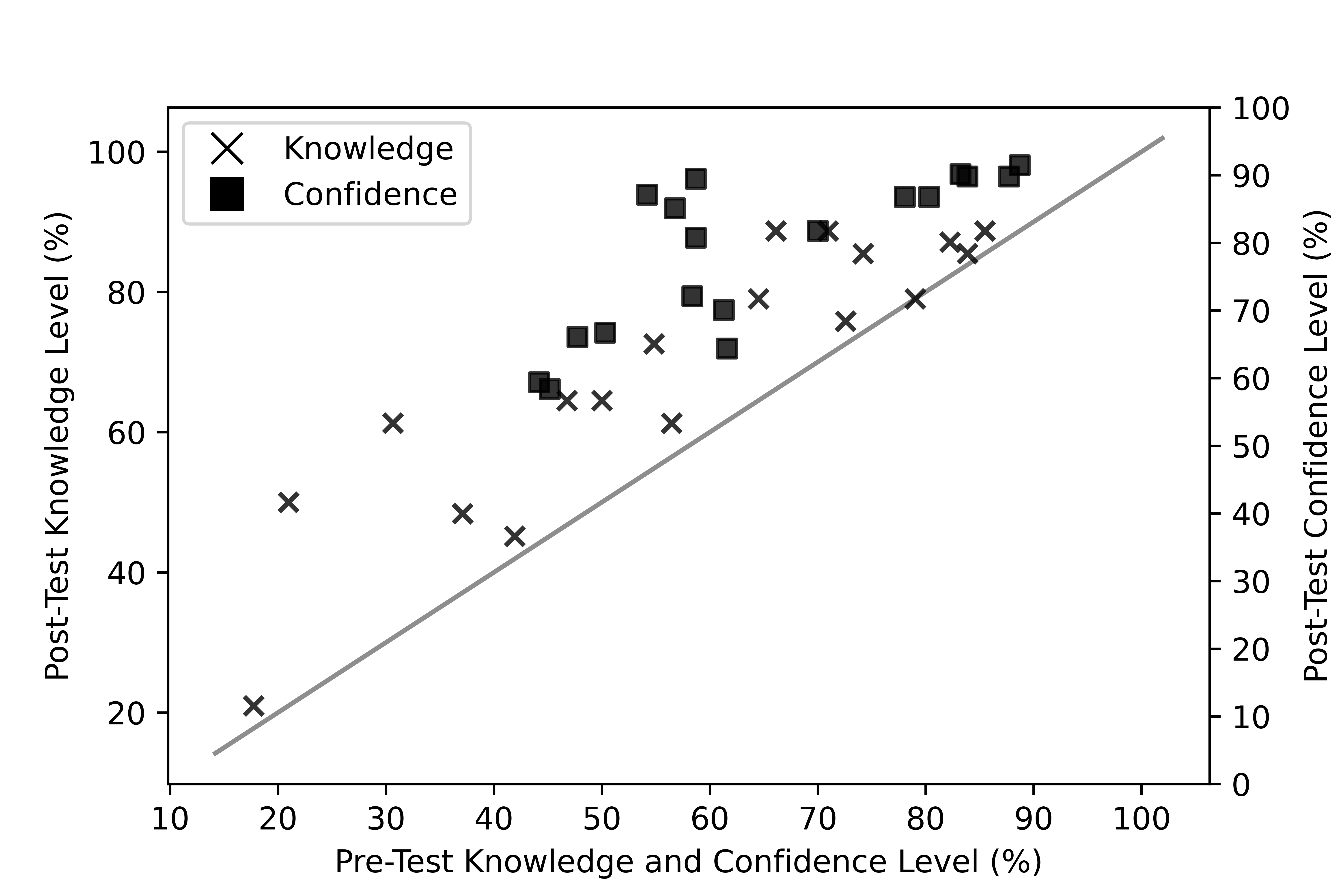}
\caption{\label{PrePost2016}Linked pre- and post-assessment knowledge (crosses) and confidence (squares) scores for participants enrolled in MIPEP 2016. The solid line marks the boundary where no knowledge or confidence gain took place. Data on the upper-left portion of the graph represent increase, while data on the lower-right portion of the graph would represent decrease.}
\end{figure}

In this section we present results based on the pre- and post-program assessment, which was introduced as a diagnostic tool during the program. A comparison of pre- versus post-program scores for all participants from the 2016 cohort are shown in Figure \ref{PrePost2016}. Data points above the line demonstrate a gain experienced to either knowledge or confidence by a participant during the program. Average participant knowledge gain, with standard deviation and standard error, was [$K_{avg}$=0.28, SD$_K$=0.19, SE$_K$=0.04], and average participant confidence gain was [$C_{avg}$=0.63, SD$_C$=0.19, SE$_C$=0.04]. With one exception, all teachers demonstrated at least some gain to their physics knowledge during MIPEP, while all teachers experienced an increase in confidence for their answers.

\begin{figure}
\subfloat{%
  \includegraphics[width=.48\textwidth]{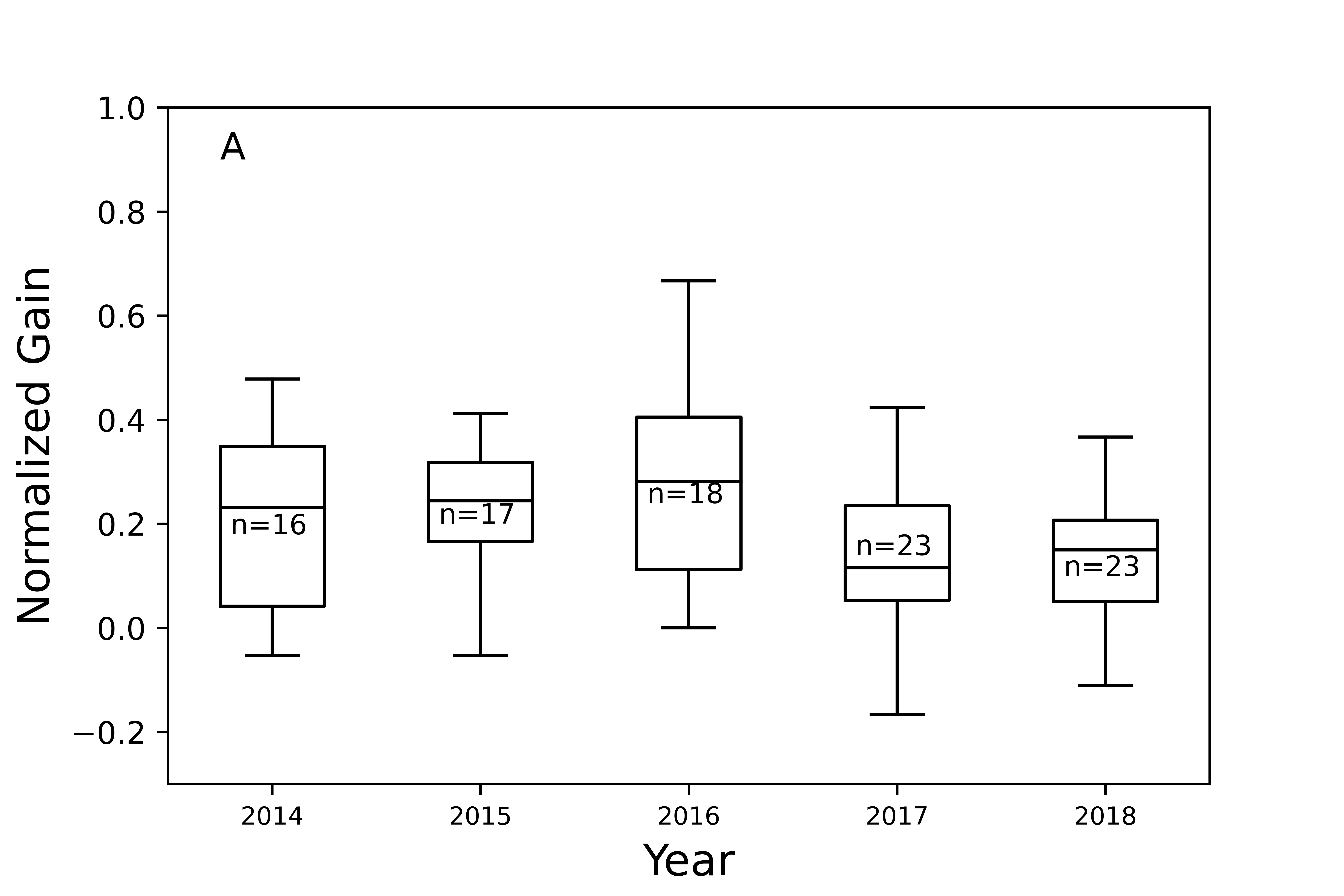}%
}\hfill
\subfloat{%
  \includegraphics[width=.48\textwidth]{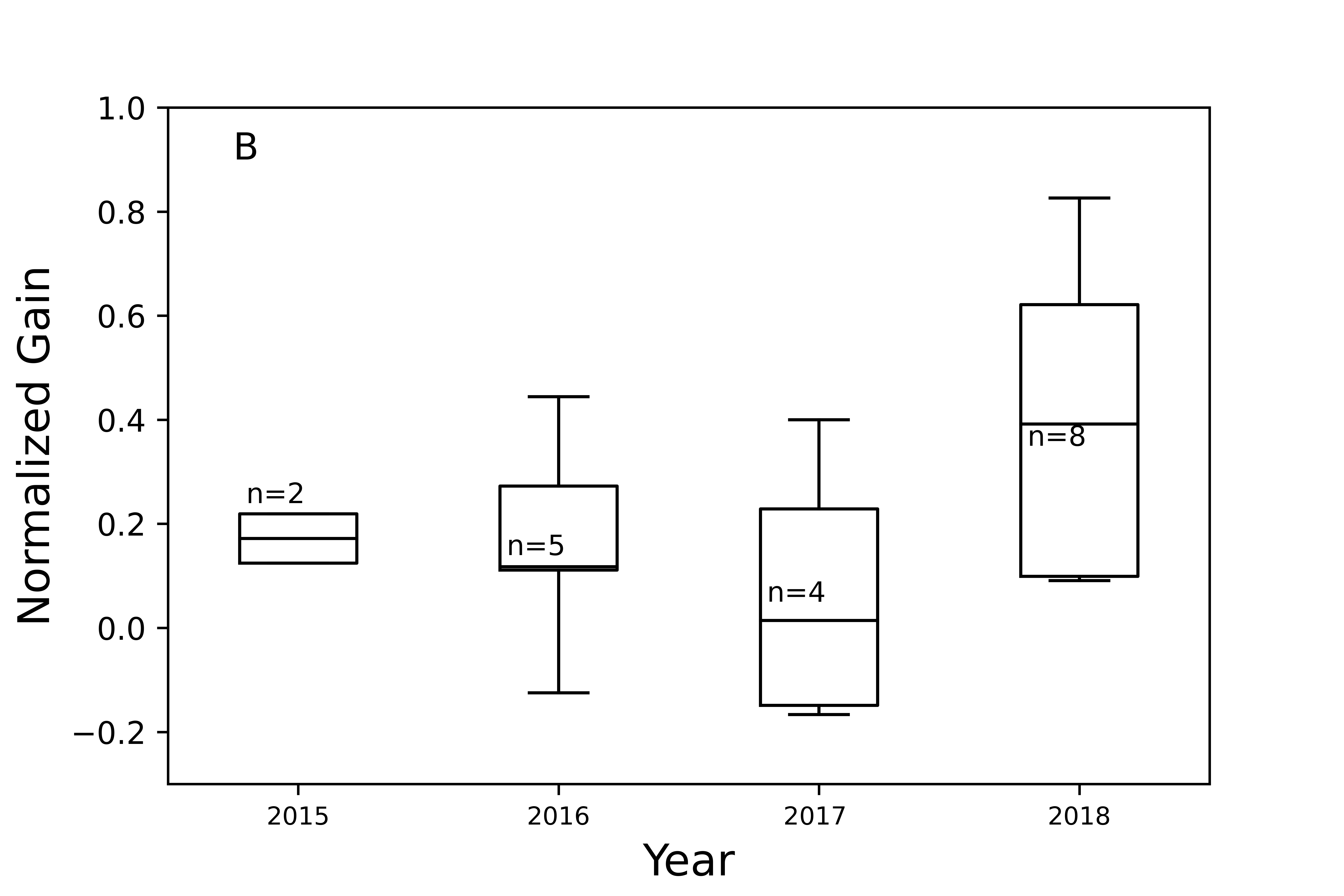}%
}
\caption{Normalized gain for teacher knowledge between (A) the pre- and post-test and (B) the pre- and subsequent testing in fall 2018. The lines in the box represent the median, the boxes represent the two middle quartiles, and the error bars represent the highest and lowest quartiles. No responses to the subsequent testing were received for participants from 2014.}
\label{Knowledge20142018}
\end{figure}

\begin{figure}
\subfloat{%
  \includegraphics[width=.48\textwidth]{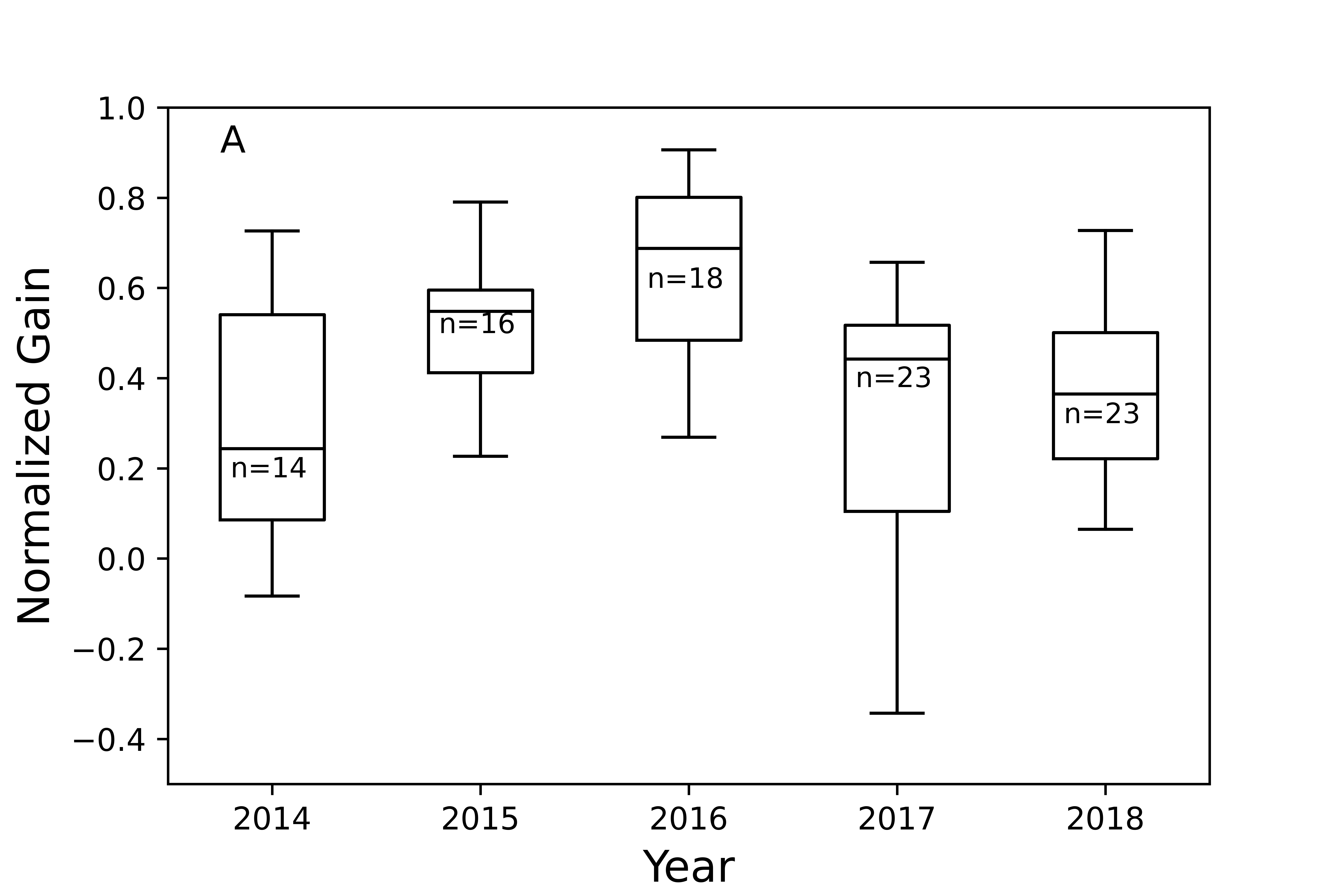}%
}\hfill
\subfloat{%
  \includegraphics[width=.48\textwidth]{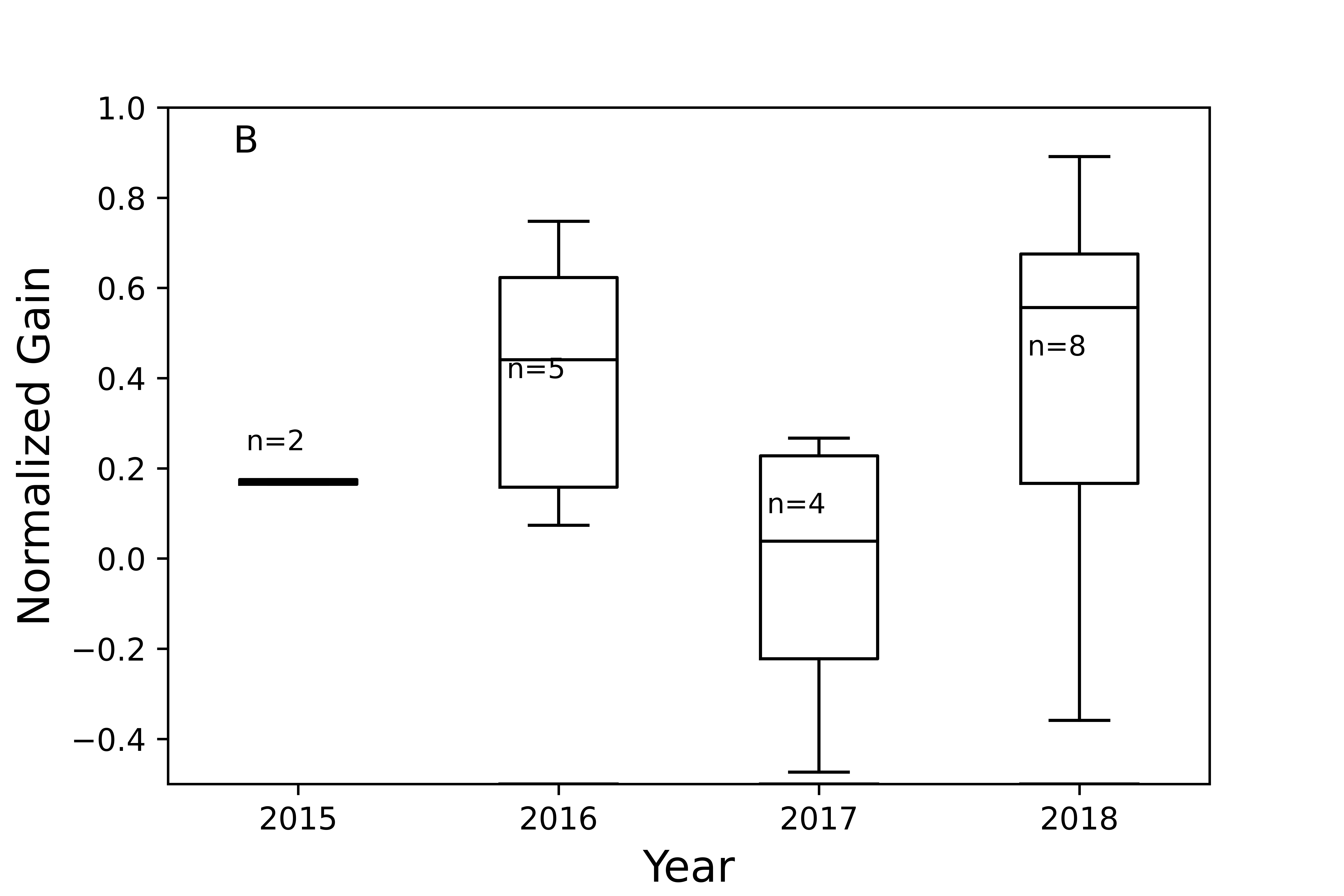}%
}
\caption{Normalized gain for teacher confidence between (A) the pre- and post-test and (B) the pre- and subsequent testing in fall 2018. The lines in the box represent the median, the boxes represent the two middle quartiles, and the error bars represent the highest and lowest quartiles. No responses to the subsequent testing were received for participants from 2014.}
\label{Confidence20142018}
\end{figure}

Normalized gains for participant knowledge and confidence for the 2014-2018 cohorts are shown in Figures \ref{Knowledge20142018} and \ref{Confidence20142018}. The normalized gains are shown between pre- and post-assessment as well as between pre- and the subsequent assessment. Results are shown only for participants who had matched data between the two assessments. Though individual gains varied, the majority of participants exhibited a positive gain to their physics knowledge and confidence over the two weeks of the program.

\begin{figure}
\centering
\includegraphics[width=8.6cm]{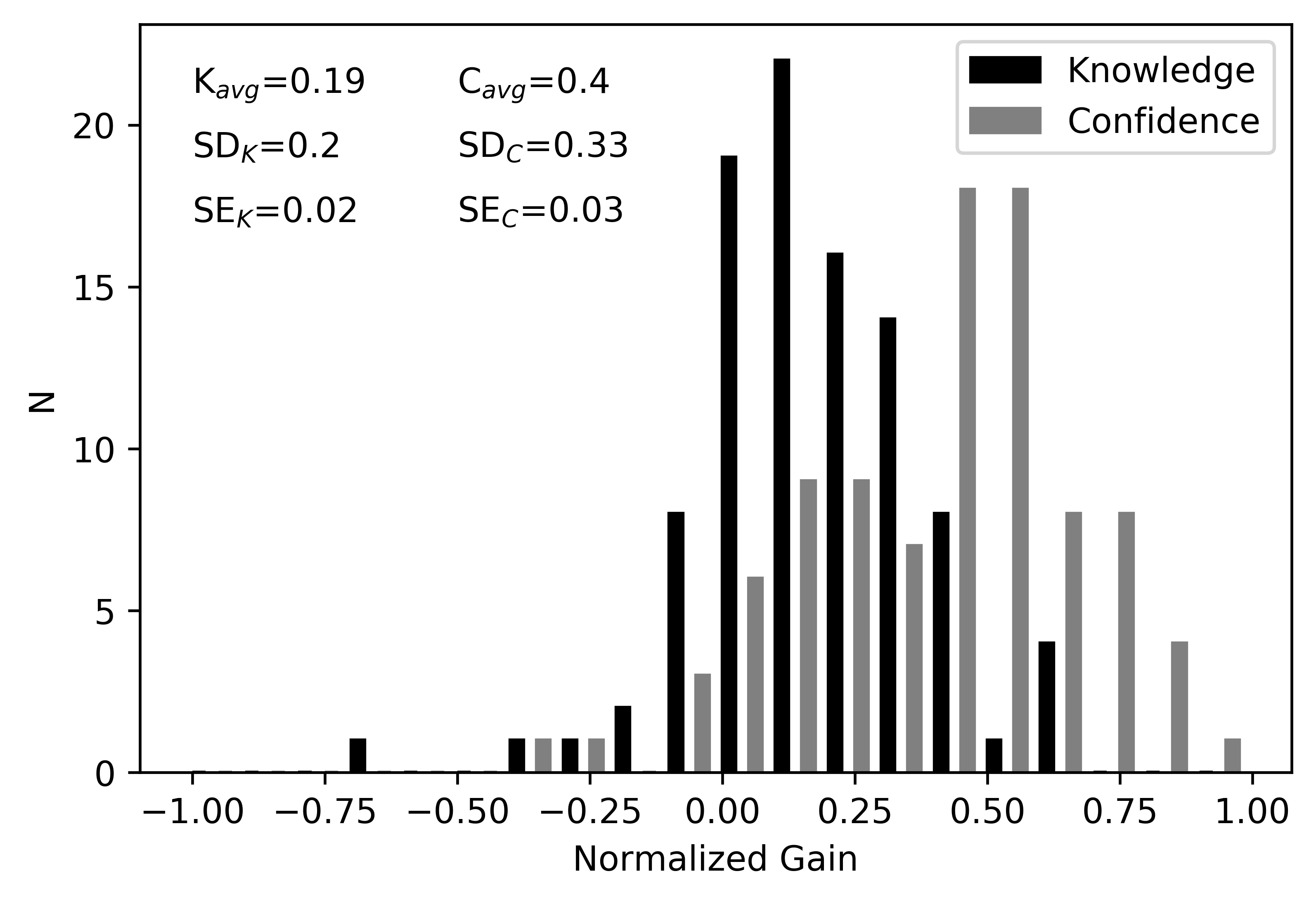}
\caption{\label{GainHist_PrePost} Distribution of normalized gain between pre- and post-assessment for participant knowledge (K) and confidence (C). The average (avg), standard deviation (SD), and standard error (SE) for each distribution have also been included.}
\end{figure}

\begin{figure}
\centering
\includegraphics[width=8.6cm]{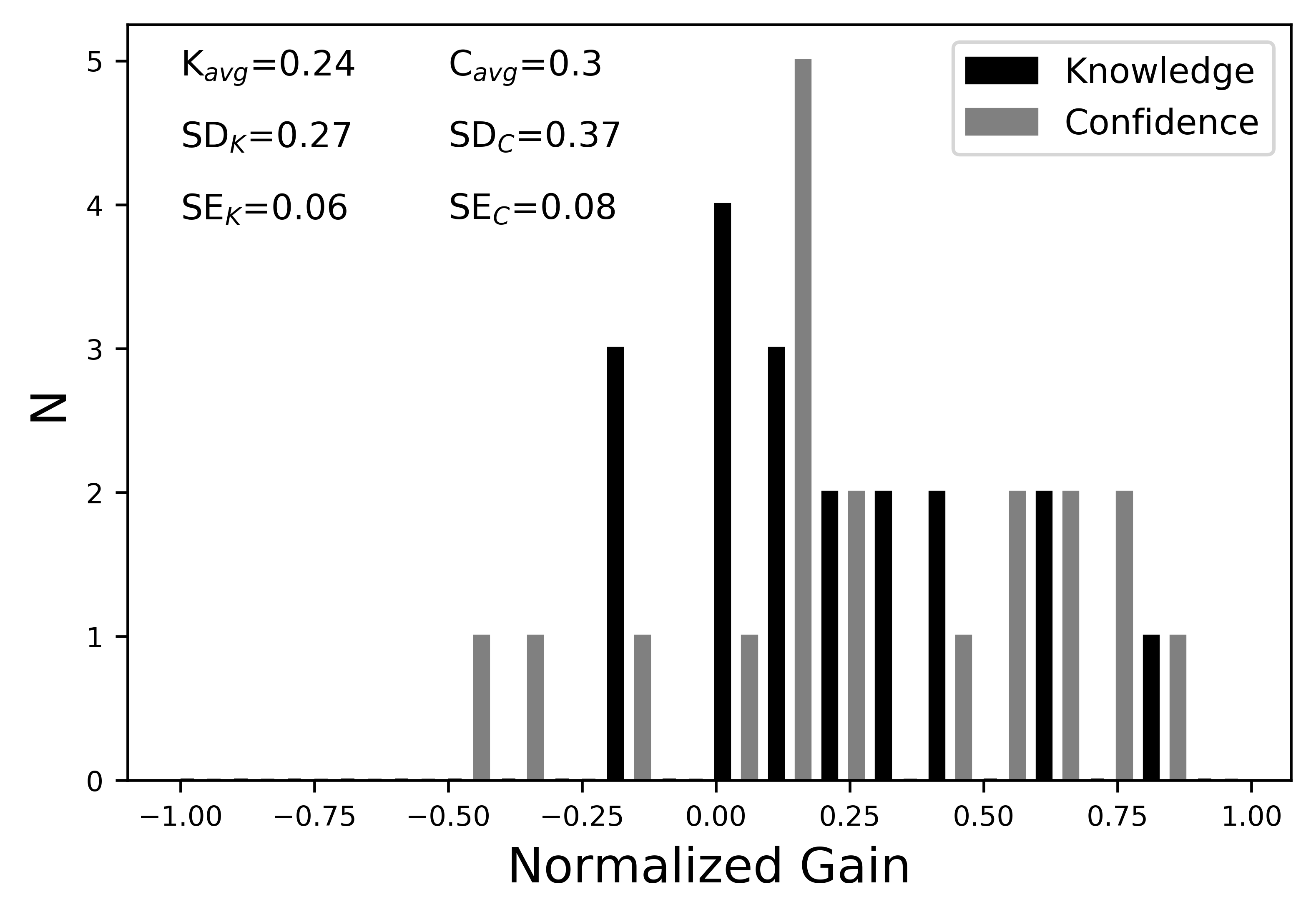}
\caption{\label{GainHist_PreSub} Distribution of normalized gain between pre- and subsequent assessment for participant knowledge (K) and confidence (C). The average (avg), standard deviation (SD), and standard error (SE) for each distribution have also been included.}
\end{figure}

Distributions of the normalized gain for knowledge and confidence for all years combined are shown in Figures \ref{GainHist_PrePost} and \ref{GainHist_PreSub}. In Figure \ref{GainHist_PrePost}, there were $N=97$ matched assessments for knowledge, and $N=94$ matched assessments for confidence. In Figure \ref{GainHist_PreSub} there were $N=19$ matched assessments for both knowledge and confidence. For all years combined, there was an average knowledge gain [$K_{avg}$=0.19, SD$_K$=0.20, SE$_K$=0.02] and an average confidence gain of [$C_{avg}$=0.40, SD$_C$=0.33, SE$_C$=0.03] between the pre- and post-assessments, Figure \ref{GainHist_PrePost}. Data from the subsequent assessment show that there was an average knowledge gain of [$K_{avg}$=0.24, SD$_K$=0.27, SE$_K$=0.06] and an average confidence gain of [$C_{avg}$=0.30, SD$_C$=0.37, SE$_C$=0.08] between the pre- and subsequent assessments, Figure \ref{GainHist_PreSub}. For all years participants maintained knowledge gains compared to their pre-program levels, with the exception of 2017 where the mean was close to zero. This cohort had relatively modest gains during the program as well. It should be noted that responses to the subsequent assessment sent in fall 2018 were received from a small number of former MIPEP participants. There are gains in participant confidence as measured by the subsequent assessment, Figure \ref{Confidence20142018}, albeit more modest in comparison to immediately post program.

\begin{figure}
\centering
\includegraphics[width=8.6cm]{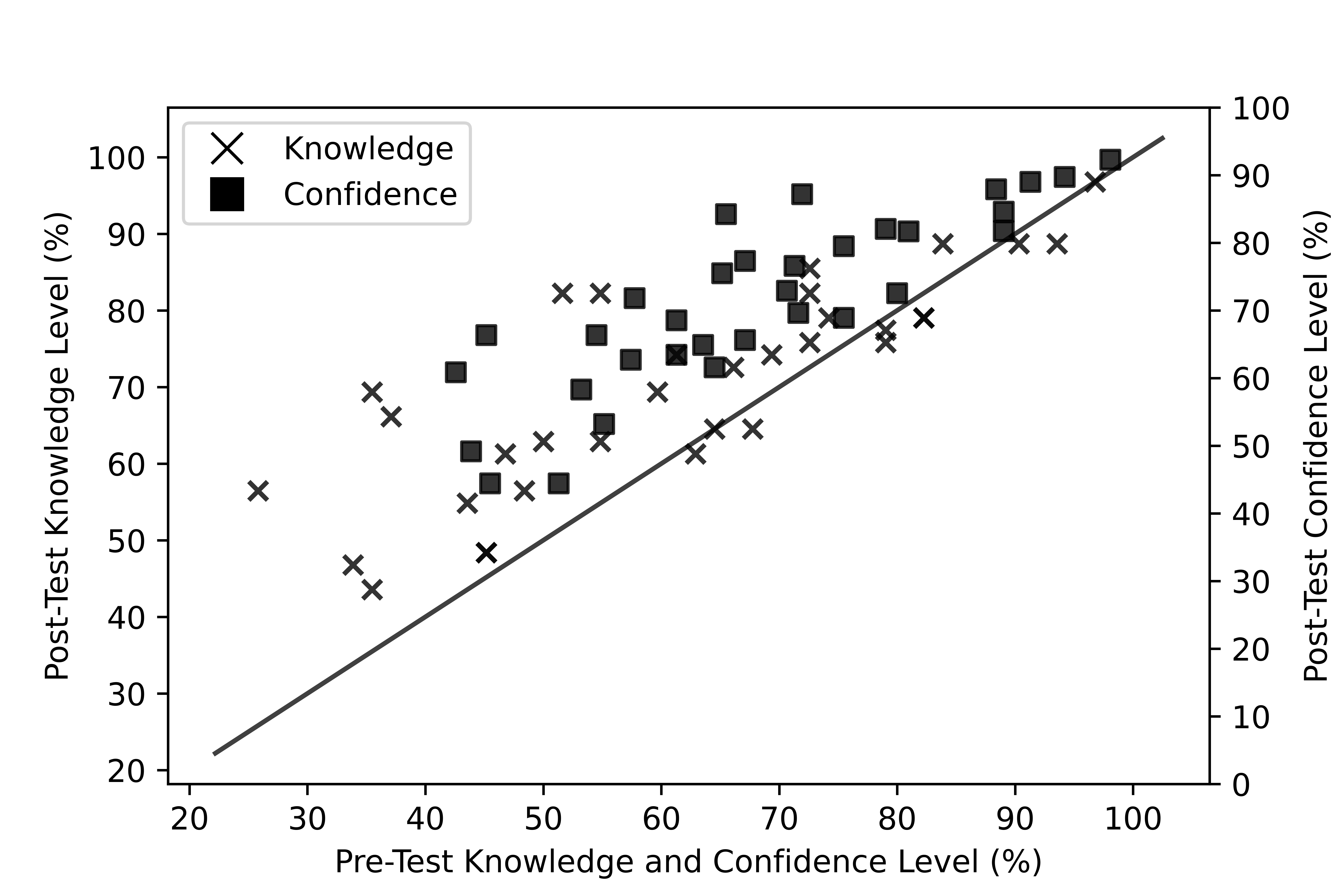}
\caption{\label{2020scatter} Linked pre- and post-test knowledge (crosses) and confidence (squares) scores for in-service teachers enrolled in MIPEP 2020. The solid line marks the boundary where no knowledge or confidence gain took place. Data on the upper-left portion of the graph represent increase, while data on the lower-right portion of the graph would represent decrease.}
\end{figure}

\begin{figure}
\centering
\includegraphics[width=8.6cm]{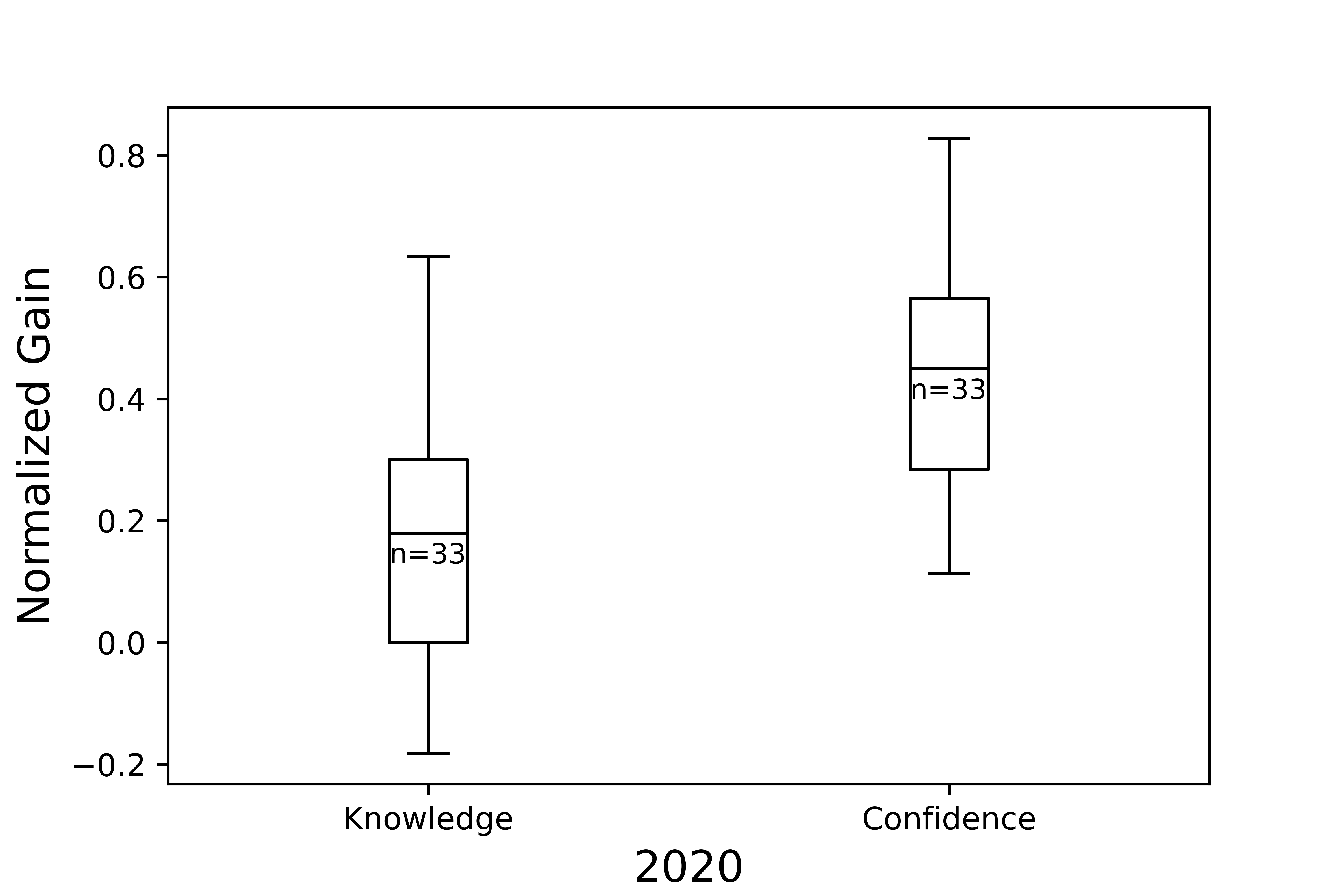}
\caption{\label{gains2020} Normalized gain for teacher knowledge and confidence between the pre- and post-test for MIPEP 2020. The lines in the box represent the median, the boxes represent the two middle quartiles, and the error bars represent the highest and lowest quartiles.}
\end{figure}

Individual participant pre- and post-assessment knowledge and confidence for MIPEP 2020 are shown in Figure \ref{2020scatter}, and participant knowledge and confidence gains are shown in Figure \ref{gains2020}. A majority of participants exhibited small gains to their knowledge, while confidence improved during the program for all. The normalized gains are shown between pre- and post-testing, which incorporate the $N=33$ paired responses for this year of the program. The average knowledge gain for the 2020 cohort was [$K_{avg}$=0.15, SD$_K$=0.27, SE$_K$=0.05] and the average confidence gain was [$C_{avg}$=0.43, SD$_C$=0.20, SE$_C$=0.03]. Overall, this cohort had modest gains for both knowledge and confidence during the program compared to years past. 

\section{Discussion}
High school enrollment in physics has been steadily increasing for decades, with a sharp increase for Texas students in the early 2000's \cite{Yoon2017}. Yet less than 40\% of high school physics teachers have a background in the field \cite{White&Tyler2014}. Prior literature has emphasized the importance of teachers receiving an appropriate depth of physics knowledge to be an effective teacher; a depth best gained through interactions with a university or college physics department \cite{McDermott1990, Foote2018}. These interactions can help teachers form their physics pedagogical content knowledge, an essential combination of subject knowledge and instructional strategies that is unique to teachers within the field \cite{Etkina2010, Magnusson1999}. The Mitchell Institute Physics Enhancement Program (MIPEP) was created as one measure to help address the need for better prepared high school physics instruction by targeting in-service teachers who have a limited background in the subject. The goal of MIPEP is to enhance participant knowledge of both physics and the relevant pedagogies which effectively support the learning of physics. Over two weeks each summer, cohorts of MIPEP participants receive intensive instruction on curriculum aligned with state standards (TEKS) which are complemented by hands-on activities and discussions with master teachers focused on instructional strategies. Physics knowledge and confidence were sampled at the beginning and end of MIPEP through a diagnostic 62-item assessment administered by master teachers working with the program. Participant performance on the assessment had no impact on their completion of the program. In fall 2018, a subsequent assessment was sent to prior participants from the 2012-2018 cohorts to provide an additional measure on the persistence of their knowledge and confidence gains. 

Improvements to participant content knowledge and confidence were measured using normalized gain. Knowledge was measured in a binary fashion, while confidence was reported using a 5-point Likert scale for each item. Though individual gains varied from teacher to teacher, results from the previous section demonstrate a positive impact on participants during the two week program. The increase in participants' confidence in their responses may indicate a boost to their physics self-efficacy \cite{Guskey1988} which may result in more effective instruction. The role of MIPEP as a professional development program can then be inferred to be successful in delivering essential information to support more effective physics instruction at the high school level.

To positively impact high school physics classrooms across Texas, professional development programs such as MIPEP must promote lasting gains to teachers knowledge, confidence, or instructional abilities. Our results, shown in Figures \ref{Knowledge20142018} - \ref{GainHist_PreSub}, indicate persistent knowledge and confidence gains for MIPEP participants. At the time of the subsequent assessment, there were both knowledge and confidence improvements for the most recent cohort. This is unsurprising as this cohort of teachers was guaranteed to be teaching physics that year, and they were in the middle of teaching material for the first time after the program. For the 2015-2017 cohorts, participant gains declined since their participation in the program, though many still showed positive gains in their knowledge and confidence compared to when they began MIPEP. This persistence of gains mirrored the impact of a K-12 professional development program run for Ohio math and science teachers which demonstrated that gains made within the program were sustained over the next three years \cite{Supovitz2000}.

As all education programs underwent an emergency shift to remote methods in 2020, MIPEP was reformatted to a Zoom-based classroom. Other than the change in delivery method, the majority of the course remained the same as years past. Participants were sent kits with materials for hands-on learning so they could engage in those activities on their own and collaboratively through Zoom with other instructors. Online instruction made the program more accessible and reduced program costs. As per years past, MIPEP 2020 administered the same 62-item conceptual assessment, this time via Google Forms. Comparing the assessment results of MIPEP 2020 to years past, the gains to instructor knowledge and confidence were similar to the average of gains of years when there was in-person instruction. It should be noted that this cohort did not benefit from the experiential learning component of laboratory activities. It should be noted that this conversion to remote delivery was done rapidly at the end of the spring term and is unlikely to represent the potential of a virtual version of this program.

The benefits of a program like MIPEP to state high school physics education is not limited to participating instructor knowledge and confidence. MIPEP also provided participants a community of like-minded people of various experiences. Since cognition is social in nature and distributed across the individual, other persons, and tools \cite{PutnamBorko2000}, it is important for a professional development program like MIPEP to place emphasis on fostering a community for these instructors who likely experience isolation in their schools \cite{White&Tesfaye2012}. Studies show that a community for physics teachers can greatly improve professional growth \cite{NehmehKelly2018, Akerson2009}. MIPEP fostered what is defined as a well-designed community of practice: one that allowed for participation in group discussion, one-on-one conversations and expert discussions \cite{WengerMcDermottSnyder2002}. In addition to pedagogical tools that can be easily implemented in their classrooms, MIPEP also provides opportunities for teachers to extend their teaching to events outside of the classroom, such as the Texas A\&M Physics \& Engineering Festival. In the shift to an online format, more teachers can be involved in the program. This can lead to a more diverse group, which has been found to lead to more effective communities of practice \cite{Keown2009}. At the same time, the in-person boarding-school format with teachers spending all the time together as a group is likely to be more effective in forming a closely knit community and camaraderie between the participants as compared to remote instruction.

\section{Limitations}
While MIPEP has been seen to have a positive impact on physics knowledge and confidence for in-service high school physics teachers in Texas, we must note some limitations to this work. With cohorts between 15-40 participants per year, MIPEP impacts a small number of high school teachers who are motivated enough to self-select into this type of intensive professional development program. The conclusion about the persistence of knowledge and confidence gains can only be held with modest certainty due to the low response rate, particularly for the oldest cohort. This low participation on the subsequent assessment is possibly due to the significant time investment requested from prior participants. Results from the 2020 cohort do not necessarily represent the potential of the program's virtual delivery as the shift from in-person to online was done rapidly due to the COVID-19 pandemic.

\section{Conclusion}
We have reported on the design and implementation of a professional development program targeted towards in-service high school physics teachers in Texas who have a limited background in the subject. Through a two-week program, participants receive intensive physics content instruction, participate in hands-on experiments, and receive instruction on research-based pedagogical strategies. Traditionally, MIPEP seeks to improve upon participating teachers' knowledge and confidence through a face-to-face program, though the program was shifted to a virtual format in 2020 due to the COVID-19 pandemic. Analysis of a diagnostic assessment administered at the beginning and end of the program demonstrated that the majority of participants increased their physics content knowledge as well as their confidence during both traditional and virtual program deliveries. The persistence of these gains was measured by distribution of a subsequent assessment sent in fall 2018. By design, MIPEP seeks to reduce isolation in remote Texas schools by providing a community of practice for high school physics teachers. We have shared the structure of this program so that it may potentially serve as a model for other institutions looking to close the preparation gap for high school instructors in their area. If expanded or replicated, MIPEP or a similar program could broaden the educational impact to high school physics education, both through improved instruction for teachers and by establishing a dialogue between those teachers and university educators.

\begin{acknowledgments}
We are grateful to the MIPEP master teachers Paula Hiltibidal and Dr. Kris Tesh for their support, providing data, and useful discussions. MIPEP exists due to generous support of the Mitchell Foundation. We also thank Dr. Jacqueline Stillisano for her contributions to the data used for this study, as well as Dr. Bhaskar Dutta and Dr. Alexey Belyanin for their leadership of MIPEP and helpful discussions. 

\end{acknowledgments}

\bibliography{bibliography}

\end{document}